\documentclass[aps,pre,reprint,amsmath,amssymb,graphicx,longbibliography,superscriptaddress]{revtex4-1}

\usepackage{bm}

\usepackage{indentfirst}
\usepackage{psfrag}
\usepackage{epsfig}
\usepackage{color}

\begin{document}

\title{Contribution  of evanescent waves to the effective medium of disordered waveguides}

\author{Miztli Y\'epez}
\affiliation{Condensed Matter Physics Center (IFIMAC), Departamento de F\'isica de la Materia Condensada and Instituto ``Nicol\'as Cabrera'', Universidad Aut\'onoma de Madrid, E-28049, Madrid, Spain.}
\affiliation{{Departamento de F\'isica, Universidad Aut\'onoma Metropolitana-Iztapalapa, Apartado Postal 55-534, 09340 M\'exico Distrito Federal, M\'exico.}
}

\author{Juan  Jos\'e S\'aenz}
\affiliation{Condensed Matter Physics Center (IFIMAC), Departamento de F\'isica de la Materia Condensada and Instituto ``Nicol\'as Cabrera'', Universidad Aut\'onoma de Madrid, E-28049, Madrid, Spain.}
\affiliation{Donostia International Physics Center (DIPC), Paseo Manuel Lardizabal 4, 20018 Donostia-San Sebastian,
Spain.}

\begin{abstract}

We consider a wave propagating through a thin disordered slab inside a wire or waveguide of finite width. In the dense weak scattering limit, the statistics for the  {\em complex reflection and transmission coefficients} (the coherent field) is found to depend dramatically on the contribution of evanescent modes or closed channels, leading to an effective refractive index whose real part is quite sensitive to the closed channels inclusion. In contrast, evanescent modes play no role in the statistical average of {\em transmittances and reflectances}. The theoretical predictions, based on the perturbative Born series expansion, are in excellent agreement with numerical simulations in disordered wires.

\end{abstract}


\date{\today}

\maketitle

\tableofcontents{}

\section{Introduction}

The coherent transport of electromagnetic, electronic and acoustic waves through random media has long been a central issue in Physics \cite{AIshimaru:1978}. In recent years, there has been a revival of interest in the impact of inhomogeneities and defects in the transmission properties in quasi-one-dimensional (Q1D) systems \cite{Mello:2010} and is the subject of intense research including  the electronic conductance of nanowires, nanotubes and nano ribbons \cite{Markussen:2007,Roche:2008,Feilhauer:2011}, the propagation of  slow light through disordered photonic-crystal waveguides \cite{Patterson:2009,Thomas:2009,Mazoyer:2010,Ott:2010} or acoustic propagation \cite{Tomsovic:2012}.

A correct description of wave transport through inhomogeneous media involves not only ``propagating'' traveling modes (``open'' channels) but it is also very sensitive to near-field ``evanescent'' modes (``closed'' channels) \cite{Carminati:2000,Graphene}. The contribution of evanescent modes has been known for a long time in the analysis of inhomogeneous waveguides and  periodic structures like multilayered frequency selective surfaces \cite{Mittra}, being particularly important near the thresholds of new propagating modes where the coupling between open and closed channels lead to geometric resonance effects \cite{Bagwell:1990,Kunze:1992,PRB53:1996_15914}. However, the statistical properties of transport through  disordered Q1D systems, obtained from numerical calculations, do not depend on the evanescent modes \cite{Froufe:2007,Payne:2013} and they are in good agreement with theoretical scaling approaches \cite{Mello:2010,DMPK,nonlinearsigma,DMPK2,PRB46:1992,PRB37:1988,Froufe:2007}, including the celebrated Dorokhov-Mello-Pereyra-Kumar (DMPK)  \cite{DMPK} and nonlinear sigma-model \cite{nonlinearsigma} approaches, which do not take into account explicitly evanescent modes. Why evanescent modes are critical for a given configuration while it seems they play no role after statistical averaging over different configurations? One of our main goals here is to provide an  answer to this open question.

In this Letter, we use numerical simulations and a perturbative analytical approach to analyze the influence of evanescent modes on the statistical properties of the transport coefficients of a thin disordered slab inside a wire or waveguide. The influence of the evanescent modes is analyzed in the asymptotic region for the far field and no near to the physical interface of the disordered region, where is well known that the evanescent modes affect strongly the statistics of the near-field \cite{Carminati:2000}. The wave number $k$ is considered halfway between the threshold of the last propagating mode and the first evanescent mode; this avoids the strong influence of the first evanescent mode on the statistical scattering properties when $k$ is near to the threshold of a new propagating mode \cite{Bagwell:1990}.

Assuming a weak scattering, non-dissipative, random medium with white-noise statistics, we find that the statistical averages of (power) transmittances and reflectances are solely determined by the mean free paths (MFPs) (in agreement with scaling theory). As we will show, in the so-called dense weak scattering limit (DWSL) \cite{PRB46:1992,Froufe:2007}, the MFPs themselves do not depend on the evanescent modes. In striking contrast, the statistical average of the scattered field (coherent field) present a completely different behavior (which was not captured by  previous scaling approaches). The propagation of the coherent wave field \cite{Foldy} is characterized by an effective wave number, whose real part determines the speed of propagation, while its imaginary part represents the losses due to scattering (often known as waveguide extrinsic losses). As the wave propagates, the amplitude of the coherent part decays exponentially, whose rate decrease is the scattering MFP. We will see that the scattering MFP is insensitive to evanescent modes, while changes in the phase of the coherent field (i.e. in the  real part of the effective wave number) are solely related to evanescent modes.

\section{Model system and numerical results}

\begin{figure}[t]
\begin{center}
\includegraphics[width=8cm]{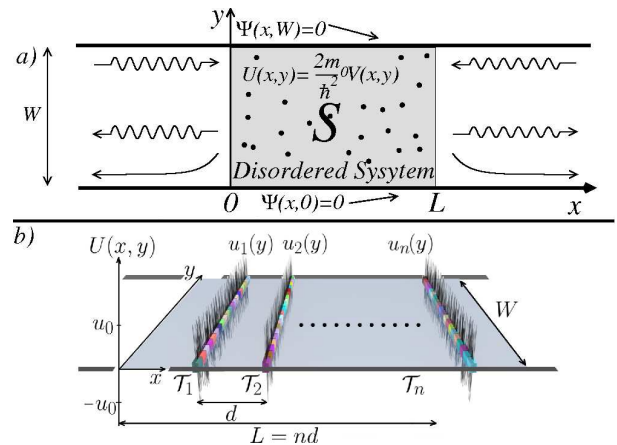}
\caption{\small{(Color online) (a) Sketch of the disordered slab in a waveguide. The fluctuating potential  is considered as a sequence of thin random slices (b) with   zero mean scattering potential (for electromagnetic waves, this would correspond to random inclusions with permittivity slightly larger and lower than the background permittivity).}}
\label{sketch} 
\end{center}
\end{figure}

Consider a wave propagating along the $x$-direction in the two-dimensional (2D) waveguide sketched in Fig. \ref{sketch}. In this work we restrict ourselves to the problem of scalar waves following the equation $\bm{\nabla}^2 \Psi + k^2 \Psi = U \Psi $. For electron transport, $\Psi$ and $U$ represent  the wave-function and the random potential ($U(\rho)= 2 mV(\rho)/\hbar^2$), respectively and $k=2\pi/\lambda$ is the wave number in absence of fluctuation ($k^2 = 2 mE_F/\hbar^2$ for free electrons with effective mass $m$ and Fermi energy, $E_F$). For electromagnetic waves, $\Psi$ would be the electric field normal to the ($x,y$) plane propagating through a waveguide filled with a material whose refractive index  fluctuates around an averaged value, e.g. a fluid with spatial density fluctuations or a dielectric with real and positive relative permittivity $\epsilon_h$ (in general, $\epsilon_h > 1$) and  random inclusions  with fluctuating permittivity $\Delta \epsilon(x,y)$ in the region $0\le x \le L$. $U$ would then be given by $U(\rho)=-k^2 \Delta\epsilon(\rho)/\epsilon_h$ with $k=(\omega/c)\sqrt{\epsilon_h}$ ($\omega$ the frequency and $c$ the speed of light).  The lateral confinement define a set of waveguide eigenfunctions

\begin{equation}
\phi_{\pm b} (\rho)  \equiv \chi_b(y) \varphi_{\pm b}(x)
 = \sqrt{\frac{2}{W}} \sin \left(\frac{b\pi}{W}y\right) \frac{e^{\pm ik_b x}}{\sqrt{k_b}},
\end{equation}

\noindent where the integer number $b$ labels the wave modes, also referred to as scattering channels. Modes with $b \le N$ ($\le kW/\pi$), are propagating modes with longitudinal wave number $k_b=\sqrt{k^2-(b\pi/W)^2}$ real; {  {$N$ denotes the number of traveling modes supported by the waveguide.}} Those modes with $b>N$ have an imaginary wave number and represent evanescent modes.

The disordered slab is constructed as a sequence of $n$ ($\gg 1$) statistically independent and identically distributed  thin scattering units of thickness $\delta$, with $k\delta \ll 1$. The scattering units are separated one each other by a fixed distance $d \gg \delta$ in the wave propagation direction $x$, with  $L=nd+\delta$ as it is sketched in Fig. \ref{sketch}. The {\it r-th} scattering unit ($r=0,1,2,\cdots ,n$), centered at $x_{r}=rd$, is specified by its potential

\begin{equation}
U_r(\rho) =  u_r(y) \Theta\left(\frac{\delta}{2}-\left|x-x_r\right|\right).
\end{equation} 

\noindent $u_r(y)$ is again constructed as a sequence of $m$ ($= W/\delta \gg 1$) segments of length $\delta$ centered at {  {$y_s= \left(s-1/2\right)\delta$}}, 

\begin{equation} 
u_r(y)=\sum_{s=1}^m u_{rs} \Theta\left(\frac{\delta}{2}-|y-y_s|\right),
\label{uy}
\end{equation}

\noindent where the potentials in each scattering unit $u_{rs}$ are statistically independent and uniformly distributed between $-\sqrt{3}u_0$ and $+\sqrt{3}u_0$, i.e. with
\begin{equation} 
\langle u_{rs}   \rangle = 0 \quad,	\quad \langle u_{rs} u_{r's'}  \rangle = u_0^2 \delta_{ss'}\delta_{rr'}.
\label{statuy}
\end{equation}

\noindent The angle brackets represent the ensemble average over different disorder realizations. If the limits $\delta \rightarrow 0$ and $d \rightarrow 0$ are considered, our model approaches the standard Gaussian white noise-like potential models.
 
For each microscopic realization of disordered slab, the (Fresnel) complex transmission, $t_{aa_0}$, and reflection, $r_{aa_0}$, coefficients (for a wave incident in channel $a_0$ and scattered in channel $a$) are numerically obtained by using a generalized scattering matrix (GSM) technique see Refs. \cite{Mittra,Froufe:2007,Torres:2004} and App. \ref{GSM_Method}. We also obtain the corresponding  (flux/power) transmittances and reflectances $T_{aa_0} \equiv |t_{aa_0}|^2 $ and $R_{aa_0} \equiv |r_{aa_0}|^2$, the total channel transmittance  $T_{a_0} = \sum_{a=1}^N T_{aa_0}$, total reflectance  $R_{a_0} = \sum_{a=1}^N R_{aa_0}$ as well as the  dimensionless conductance of the waveguide, $g \equiv \sum_{a_0=1}^N T_{a_0} $. Although the transport coefficients link (far-field) incoming and outgoing propagating channels, their actual values  are known to be strongly dependent on the number of closed channels $N^{\prime}$ (in practice, $N^{\prime}$ is increased until convergence of the numerical results). Independently of $N^{\prime}$ the GSM approach  guaranties  flux conservation ($T_{a_{0}}+R_{a_{0}} =1$) for each realization.

\section{Evanescent modes and statistical averages}

\begin{figure}[t]
\begin{center}
\includegraphics[width=8.5cm]{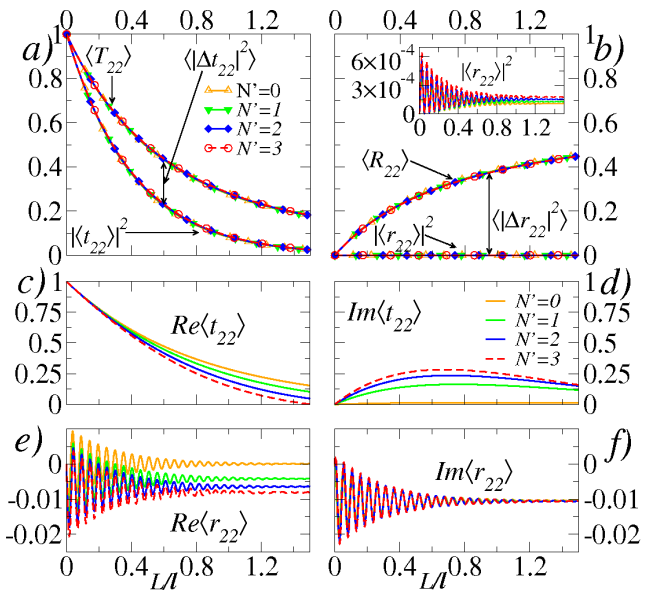}
\end{center}
\caption{\small{
{{
(Color online) Numerical results. $a$-b) Transmittance $\left\langle T_{22}\right\rangle$, Reflectance $\left\langle R_{22}\right\rangle $  and their coherent intensities $\vert\left\langle t_{22}\right\rangle\vert^{2}$, $\vert\left\langle r_{22}\right\rangle\vert^{2}$. c-f) Complex Coefficients $\left\langle t_{22}\right\rangle $ and $\left\langle r_{22}\right\rangle$. Each simulation considers $N=2$ propagating modes and different evanescent modes ($N^{\prime}=0,1,2,3$).}}
}}
\label{t22} 
\end{figure}

In order to illustrate the effect of the evanescent modes on the statistical averages of different quantities, we consider a waveguide that supports $N=2$ propagating modes, with $kW/\pi = 2.5$ far away from a new propagating mode; in this case, the scattering matrix of the disordered system

\begin{equation}
S=\left(
\begin{array}{cc}
r & t^{\prime}
\\
t & r^{\prime}
\end{array}
\right),
\end{equation}

\noindent is a $4\times 4$ matrix, while its transmission $t$ and reflection $r$ blocks are $2\times 2$ matrices themselves. For this system, we perform four numerical simulations, each one considering a different number of evanescent modes $N^{\prime}=0,1,2,3$. The results for $a_0=a=2$ are summarized in Fig. \ref{t22} (a complete summary for all the other coefficients can be found in App. \ref{Complete_Num_Results}), where the statistical averages are plotted as a function of the slab thickness $L$ in units of the mean free path $\ell$. The mean free path $\ell$, is numerically obtained ($ k\ell \approx 100$) from the slope of the conductance curve at $L=0$ ($\langle g \rangle_{L}\approx N(1-L/\ell + \cdots)$) see Ref. \cite{Feilhauer:2011} and App. \ref{Borns_Details}. All the expectation values involve an ensemble average of $10^6$ different realizations of the microscopic random potential with $kd=0.1$, $k\delta=0.001$ and $u_0\delta^2 = k\delta/(4\sqrt{3})$. Our results show that the  ensemble averages of both transmittance, $\left\langle T_{aa_0} \right\rangle $ (Fig. \ref{t22}$a$) and reflectance,   $\left\langle R_{aa_0} \right\rangle $ (Fig. \ref{t22}$b$) are not sensitive to the number of closed channels that are taken into account in the calculations, a remarkable unexplained result already pointed out in previous numerical simulations for those quantities \cite{Froufe:2007,Payne:2013}. In contrast (Fig. \ref{t22}($c$-$f$)), the complex transmission and reflection coefficients are strongly dependent on the  evanescent modes field but in a peculiar way: while the real (imaginary) part of the  transmission (reflection) coefficients are rather insensitive to the number of evanescent modes (at least for small thicknesses), both $\mathrm{Im}\left\langle t_{22}\right\rangle  $ and $\mathrm{Re}\left\langle r_{22}\right\rangle $ show a clear  dependence with the number of evanescent modes. Before discussing the origin  of these, apparently, puzzling results, it is interesting to  notice that the intensity of the averaged transmitted and reflected fields (the so-called ``coherent'' intensity), defined as  $\vert\langle t_{aa_{0}} \rangle\vert^{2}$ and $\vert\langle r_{aa_{0}} \rangle\vert^{2}$, is also independent on the evanescent modes as shown in Fig. \ref{t22}$a,b$ (the difference $\left\langle T_{aa_0} \right\rangle - \vert\langle t_{aa_{0}} \rangle\vert^{2} = \langle \vert \Delta t_{aa_{0}} \vert^{2} \rangle$ corresponds to the diffuse field intensity). This suggest that evanescent modes only affect the phase of the averaged (coherent) fields: see App. \ref{Coherent_Diffsuive_Fields}.

In order to understand the numerical findings, let us first consider the transmission, $\bm{t}_r$, and reflection, $\bm{r}_r$, matrices for a single slice centered at $x=x_r$. Expanding the wave function in transverse eigenfunctions and assuming that the wave function inside the slice is constant along $x$, it is easy to obtain

\begin{equation}
\bm{t}_r = \bm{I} -\frac{i}{2} \bm{\varphi}_r^{-} \bm{\mathcal{T}}_r \bm{\varphi}_r^{+},
\quad \quad
\bm{r}_r =   -\frac{i}{2} \bm{\varphi}_r^{+}\bm{\mathcal{T}}_r \bm{\varphi}_r^{+} ,
\end{equation}

\noindent where $ \bm{\varphi}_r^{\pm}$ are diagonal matrices, $\left(\bm{\varphi}_r^{\pm}\right)_{bb'} = \varphi_{\pm b}(x_r) \delta_{bb'}$ and $\bm{\mathcal{T}}_r$ is the transition $\mathcal{T}$-matrix of a thin scattering unit

\begin{eqnarray}
\bm{\mathcal{T}}_r & =& \bm{u}_r \delta \left[ \bm{I} + \widehat{\bm{G}}
\bm{u}_r \delta \right]^{-1} ,
\label{weak}
\\
\left(u_{r}\right)_{bb'} &=& \int_0^W  \Big\{\chi_b(y) u_r(y)  \chi_{b'}(y) \Big\} dy,
\end{eqnarray}

\noindent being the matrix elements

\begin{eqnarray}
\widehat{G}_{bb'} &=& \delta_{bb'} \frac{1}{\delta} \int_{x_r-\delta/2}^{x_r+\delta/2} \left\{\frac{i}{2k_b} e^{ik_b |x-x_r|}\right\}dx  \nonumber 
\\
&=& \delta_{bb'} \frac{i }{2 k_b}
\left( 
\frac{e^{i\frac{ k_b \delta}{2}}-1}{\frac{i k_b\delta}{2}}
\right) .
\end{eqnarray}

In contrast with previous approaches based on point  or ``delta''-scatterers \cite{Bagwell:1990,PRB53:1996_15914}, our approach (with slices having a finite thickness $\delta$) converges for any number of evanescent modes and does not require any regularization scheme.
 
In absence of dissipation, flux conservation 
leads to the Q1D version of the Optical Theorem (OT) \cite{Messiah} for a single slice:
 \begin{equation}
\sum_{a=1}^N \frac{\left| (\mathcal{T}_r)_{a a_{0}}\right|^2}{4 k_a k_{a_0}} = - \frac{\mathrm{Im} \left((\mathcal{T}_r)_{a_{0}a_{0}}\right)}{2 k_{a_0}} ,
\label{OT}
\end{equation}
which relates the imaginary part of the forward scattering amplitude to the total scattered intensity.

In the limit $L\rightarrow 0$ (in the macroscopic sense, i.e. the slab still contains a macroscopic number of scatterers $n \gg 1$), we can estimate the effective wave number $k_{\mathrm{eff}}$ inside a thin slab by comparison between the average transmitted coherent field, $\langle \Psi \rangle$, to the result of an homogeneous uniform media \cite{Foldy,Jackson}

\begin{eqnarray}
\langle t_{a_0 a_0} \rangle &\approx& 1 -i \frac{1}{d} \frac{\langle (\mathcal{T}_{r})_{a_{0}a_{0}}\rangle}{2k_{a_{0}}}L 
= 1+ i \frac{k_{\mathrm{eff}}^2-k^2}{2k_{a_0}} L,
\\
k_{\mathrm{eff}}^2 &=& 
k^2-\frac{1}{d} \langle (\mathcal{T}_{r})_{a_{0}a_{0}}\rangle =
k^{2}\hat{n}_{a_{0}}^{2},
\label{Def_k_eff}
\end{eqnarray}

\noindent where $\hat{n}_{a_{0}}^{2}=1-\left\langle (\mathcal{T}_{r})_{a_{0}a_{0}}\right\rangle / k^{2} d$, defines the -mode dependent- refractive index of the traveling mode $a_{0}$.

For weak scattering units, Eq. (\ref{Def_k_eff}) leads to an effective longitudinal wave number inside the thin slab given by $(k_{\mathrm{eff}})_{a_0} \approx k_{a_0} + \Delta k_{a_0}$, being

\begin{equation} 
\Delta k_{a_0} = - \frac{ \langle (\mathcal{T}_{r})_{a_{0}a_{0}}\rangle}{2k_{a_0}d} \equiv  \frac{1}{\ell'_{a_0}}+ i \frac{1}{\ell_{a_0}}. \label{deltakeff}
\end{equation}

In Eq. (\ref{deltakeff}), we have identified $\ell_{a_0}^{-1}=\mathrm{Im}(k_{\mathrm{eff}})_{a_0}$ as the inverse of the scattering MFP for the incoming open channel $a_0$. $\ell_{a_0}$ can be related to the channel-channel mean free paths $\ell_{aa_{0}}$, which are associated with the incoherent sum of reflections from channel  $a_0$ to channel $a$, 

\begin{eqnarray}
\frac{1}{\ell_{aa_0}} &=& \frac{1}{d} \langle\left|(r_r)_{aa_0}\right|^2\rangle = \frac{1}{d}
\frac{ \left\langle \vert\left(\mathcal{T}_{r}\right)_{aa_{0}}\vert ^{2}\right\rangle }{4k_{a}k_{a_{0}}},
\label{Ind_MFP} 
\\
\frac{1}{\ell_{a_0}} &=& \sum_{a=1}^N \frac{1}{\ell_{aa_0}} =  - \frac{1}{d} \frac{ \mathrm{Im} \left( (\mathcal{T}_r)_{a_{0}a_{0}}\right)}{2k_{a_0}}.  
\end{eqnarray}

The last identity, in agreement with Eq. (\ref{deltakeff}), is a direct consequence of the OT for a single slice: see Eq. (\ref{OT}). Notice that our definition of the  scattering MFP, $\ell_{a_0}$, commonly used in the scaling theory of transport \cite{Mello:2010}, differs by a factor of 2  from that of the MFP $\ell_{a_0}^*$ of kinetic theory, i.e. $\ell_{a_{0}} = 2 \ell_{a_{0}}^*$.

\section{Weak scattering limit}

In the weak scattering limit, $\bm{\mathcal{T}}_r \approx \delta\bm{u}_r - \delta\bm{u}_r  \widehat{\bm{G}} \delta\bm{u}_r + \cdots $ (the two first terms corresponding  to the second order Born approximation \cite{PRB53:1996_15914}), so from Eq. (\ref{deltakeff}), we easily obtain

\begin{eqnarray}
\frac{1}{\ell_{a_{0}}^{\prime}}
 & \approx &
\frac{\delta^2}{d}\sum_{ b=N+1 } ^{N+N^{\prime}}
\frac{\langle \left(u_{1}\right)_{ba_{0}}^{2}\rangle}{2k_{a_{0}}}\widehat{G}_{bb},
\label{phase_scatt_mfp}
\\
\frac{1}{\ell_{a_{0}}} &=& \sum_{a=1}^N \frac{1}{\ell_{aa_0}}
\approx
\frac{\delta^2}{d} \sum_{a=1}^{N}
\frac{\langle \left(u_{1}\right)_{aa_{0}}^{2}\rangle }{4k_a k_{a_0}}.
\label{scatt_mfp}
\end{eqnarray}

Near the onset of a new propagating channel, $\widehat{G}_{bb}$ diverges and the weak scattering approximation is not longer valid. However, as we have mentioned before, the wave number has been considered halfway for the threshold from the last open channel and the first closed channel, i.e., $kW/\pi = N +1/2$. In this regime, while the scattering MFP $\ell_{a_0}^{-1}=\mathrm{Im}(k_{\mathrm{eff}})_{a_0}$ is independent on the evanescent modes, changes in $1/\ell_{a_0}^{\prime}$ come solely from the  evanescent modes. This result, valid for arbitrary transversal correlations inside each slab, is a key important outcome of the present work. If we restrict ourselves to  short range correlations (see Eqs (\ref{uy}) and (\ref{statuy})), $k_{\mathrm{eff}}$ can be rewritten as

\begin{equation}
k_{\mathrm{eff}}^2  = k^2 +
\int_{0}^{W} dy \chi^2_{a_0}(y) \int G(\rho,\rho')\Gamma(\rho,\rho') d^2\rho'
\end{equation}
 
\noindent with  $\Gamma(\rho,\rho') \equiv u^2 \Theta\left(\delta/2-|x-x'|\right)\Theta\left(\delta/2-|y-y'|\right)$,
which, in the limit $W \rightarrow \infty$, is equivalent to that 
obtained  for a homogeneous  infinite medium, with sub wavelength short range correlations  \cite{limit}.

\begin{figure}[t]
\begin{center}
\includegraphics[scale=0.3]{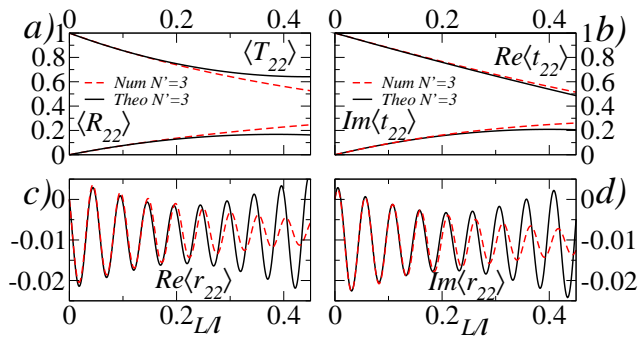}
\end{center}
\caption{\small{
{  {(Color online) Comparison between the theoretical results, Eqs. (\ref{MyPHDEq7.138})-(\ref{MPHTEq7.98}), and the numerical simulations of Fig. \ref{t22}. $a$) $\left\langle T_{22}\right\rangle$ and $\left\langle R_{22}\right\rangle$, b) $\left\langle t_{22}\right\rangle$ and c-d) $\left\langle r_{22}\right\rangle$. The comparison is done for $N=2$ and $N'=3$.}}
}} 
\label{Ret22Imt22N2Np0123} 
\end{figure}

In order to obtain a physical description of the macroscopic statistics beyond the $L\rightarrow 0$ limit, we make use of the Born series in the DWSL. Since the Born approach generates a double series expansion in powers of $L/\ell$ and in inverse powers of $k \ell$ (here $k$ denotes symbolically any possible wave number $k_b$, while $\ell$  does represent any physical parameter $\ell_{a a_0}$, $\ell_{a_0}$ or $\ell_{a_0}'$), we work in the short-wavelength approximation (SWLA) \cite{Froufe:2007}, which fix the relation between the characteristic lengths: $k\delta \ll 1 \ll kL \ll k\ell$. After a lengthly calculation, we find: 
\begin{eqnarray}
\left\langle
T_{aa_{0}}\right\rangle &=&
\left[\frac{L}{\ell
_{aa_{0}}}+\sum_{b=1}^{N}\frac{L^{2}}{\ell_{ab}\ell_{ba_{0}}}
-\left(\frac{1}{\ell_{a}}+\frac{1}{\ell_{a_{0}}}\right)\frac{L^2}{\ell_{aa_{0}}}+\cdots\right]\nonumber
\\
&&+
\delta_{aa_{0}}\left[1-2\frac{L}{\ell
_{a}}+2\left(\frac{L}{\ell_{a}}\right)^{2}+\cdots\right] + O\left(\frac{1}{k\ell}\right)\nonumber
\end{eqnarray} \begin{eqnarray}
\left\langle R_{aa_{0}}
\right\rangle &=&\biggl\{\frac{L}{\ell_{aa_{0}}}-\biggl(\frac{1}{\ell_{a}}+\frac{1}{\ell_{aa_{0}}}+\frac{1}{\ell_{a_{0}}}
\biggr)\frac{L^2}{\ell_{aa_{0}}}
 \nonumber
\\
&&+ \sum_{b=1}^{N}\frac{1+\delta_{a a_0}}{\ell_{ab}\ell_{ba_{0}}}L^2
+\cdots\biggr\}+O\left(\frac{1}{k\ell}\right),
\label{MyPHDEq7.138}
\end{eqnarray}

\noindent which shows that the transmittances and reflectances only depend on the MFP's and, as a consequence, they are independent on the number of evanescent modes. As it is shown in Fig. \ref{Ret22Imt22N2Np0123}($a$), in the ballistic regime $L < \ell$ (with $1/\ell = N^{-1}\sum_{a=1}^{N} 1/\ell_a$), there is an excellent agreement between the theoretical predictions and the numerical results of Fig. \ref{t22}($a,b$). The complex coefficients, given by 
\begin{eqnarray}
\left\langle t_{aa_{0}}\right\rangle  &=& 
\delta_{aa_{0}} \biggr[1+i\Delta k_{a} L
+\left(i\Delta k_{a}\right)^{2}
\frac{L^{2}}{2!}+\cdots \biggr]  
+O\left(\frac{1}{k\ell}\right),
\nonumber 
\\
\left\langle r_{aa_{0}}\right\rangle &=& 
\delta_{aa_{0}} \frac{\Delta k_a}{2k_{a}}\biggl\{\left(e^{2ik_{a}L}-1\right) \nonumber
\\
&&
+2 \biggl[i\Delta k_a
-\frac{1}{2\ell_{aa}}\biggr]
L e^{2ik_{a}L}
+\cdots 
\biggr\}+O\left(\frac{1}{k\ell}\right)^{2},
 \label{MPHTEq7.98} 
\end{eqnarray}

\noindent are also in excellent agreement with the numerical results as it can be seen in Fig. \ref{Ret22Imt22N2Np0123}($b-d$). The results for $\left\langle t_{aa_{0}}\right\rangle $ suggest an exponential behavior $\sim \delta_{aa_0}  e^{i\Delta k_{a} L } = \delta_{aa_0}  e^{i L/\ell'_{a_0} }e^{-L/\ell_{a_0} } $ which differs  from the simple exponential decay predicted by the scaling theory {  {given in Ref.}} \cite{PRB46:1992}. In contrast, $\left\langle r_{aa_0}\right\rangle $  oscillates  ($\propto e^{i 2 k_{a} L }$) around a  constant background $\sim - \delta_{aa_0} \Delta k_{a}/2k_{a}$ which explains the peculiar behavior of the reflection coefficients in Fig. \ref{t22}($e,f$): see details in App. \ref{Borns_Details}.

\section{Conclusions}

The numerical and theoretical results of the present work demonstrate that even in the asymptotic region and far away from the threshold of the first evanescent mode, the evanescent modes could affect the statistical scattering properties of disordered waveguides. Equations (\ref{MyPHDEq7.138}), (\ref{MPHTEq7.98}) together with (\ref{deltakeff}), (\ref{phase_scatt_mfp}) and (\ref{scatt_mfp}) summarize the main theoretical results of this Letter. They show that in the dense weak scattering limit, the mean free paths $\ell_{aa_{0}}$ do not depend on the evanescent modes which provide a simple explanation of the numerical results and the success of the scaling approach to wave transport. In contrast, we have shown that the so-called coherent field  and its effective wave number depend both on propagating  and evanescent modes which could be specially relevant to understand the influence of disorder in the propagation of slow waves in a waveguide or when dealing with a device where interferences between different wires or waveguide arms are relevant. Our predictions could be tested, for example, in microwave waveguides where analysis of transport coefficients and intensities is currently performed to obtain accurate values of the permittivity of complex solid and liquid dielectrics \cite{microwave} or by direct measurement of both real and imaginary part of the electric fields \cite{Genack}.  Equivalent measurements in the visible or infrared would require waveguide  heterodyne methods \cite{Hulst}.

\acknowledgments

We thank J. Feilhauer, L. Froufe-P\'erez and   P.A. Mello for important discussions and C. Lopez Nataren for technical support in the numerical simulations. This work has been supported by the Spanish Ministerio de Ciencia e Innovaci\'on through CSD2007-00046 (NanoLight.es); FIS2009-13430; FIS2012-36113 
and by the Comunidad de Madrid  P2009/TIC-1476. 
JJS acknowledges  an IKERBASQUE Visiting Fellowship and
M.Y. thanks the Mexican Consejo Nacional de Ciencia y Tecnolog\'ia for postdoctoral grants (No. 162768 and 187138).

\appendix
\section{Supplemental Material} 

This supplemental material gives technical support and complementary details to the information presented in the main text of the Letter. The supplemental material is organized as follows: In Sec. \ref{GSM_Method} it is introduced the concept of the generalized scattering matrix that is used to obtained the numerical results shown in the Letter. In Sec. \ref{Complete_Num_Results}, it is shown the complete set of numerical results for the expectation values of the transmittance, reflectance, conductance and the complex coefficients. In addition, the complementary information presented in Sec. \ref{Complete_Num_Results} shows that the numerical results are consistent with the flux conservation property. In Sec. \ref{Borns_Details}, we present a technical discussion of the Born series prediction given in the Letter for the expectation values of the observables on interest. Finally, in Sec. \ref{Coherent_Diffsuive_Fields} we give a brief discussion of the coherent and diffuse fields contributions.

\subsection{Generalized scattering matrix method}\label{GSM_Method}

As it is mentioned in the text, the numerical results for the expectation values of the complex reflection $\langle r_{aa_{0}} \rangle$ and transmission $\langle t_{aa_{0}} \rangle$ coefficients as well as the reflectance $\langle R_{aa_{0}} \rangle$ and transmittance $\langle T_{aa_{0}} \rangle$, are obtained by using the {\em extended} or {\em generalized scattering matrix} (GSM) technique. In this section we briefly introduce the concept of the generalized scattering matrix $\widetilde{S}$ and how this one was implemented to obtain the numerical results of the Letter. The interested reader can find details of the GSM method in Refs. \cite{Mello:2010,Torres:2004,Mittra}.

\begin{figure}[t]
\begin{center}
\includegraphics[scale=0.34]{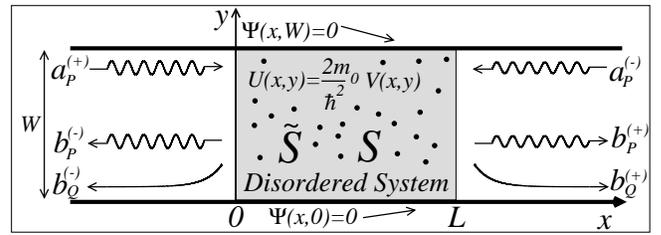}
\caption{\small{Scattering problem in a waveguide. $a_{P}^{\left(+\right)}$ and $a_{P}^{\left(-\right)}$ denote $N$ dimensional vectors, being their elements all possible incoming open channel amplitudes. Analogously $b_{P}^{\left(-\right)}$ and $b_{P}^{\left(+\right)}$ are vectors with all possible outgoing open channel amplitudes and $b_{Q}^{\left(-\right)}$, $b_{Q}^{\left(+\right)}$, are the corresponding outgoing closed channel vectors, whose dimensionality $N^{\prime}$ is, in principle, infinite.}}
\label{Theregimes} 
\end{center}
\end{figure}

In order to illustrate the implementation of the GSM technique, we first consider the most general situation of the scattering problem in a waveguide, which is shown in Fig. \ref{Theregimes}. In this figure, incoming-waves of open channels \begin{small}$a_{P}^{\left( + \right) }$\end{small} and \begin{small}$a_{P}^{ \left(- \right) }$\end{small} (incoming-waves in closed channels are not possible, so \begin{small}$a_{Q}^{\left(+\right)}=a_{Q}^{\left( -\right)}=0$\end{small}), are scattered by the disordered system, giving rise to outgoing-waves, both in open channels \begin{small}$b_{P}^{ \left(+\right) }$\end{small}, \begin{small}$b_{P}^{ \left(-\right)}$\end{small} as in closed ones \begin{small}$b_{Q}^{ \left(+ \right) }$\end{small}, \begin{small}$b_{Q}^{ \left(-\right)}$\end{small}; the symbols $+$ and $-$ denote, respectively, waves traveling to the right and to the left, while $P$ and $Q$ represent open and closed channel components, respectively. The scattering problem is formally described by the GSM $\widetilde{S}$, which relates open and closed channel outgoing-wave amplitudes to the open channels incoming-wave amplitudes, i.e., 

\begin{equation}
\left(\begin{array}{c}
b_{P}^{\left(-\right)} \\
b_{Q}^{\left(-\right)} \\
b_{P}^{\left(+\right)} \\
b_{Q}^{\left(+\right)}
\end{array}\right)
=\widetilde{S} \left(\begin{array}{c}
a_{P}^{\left(+\right)} \;\;\;\;\;\;\\
a_{Q}^{\left(+\right)}=0 \\
a_{P}^{\left(-\right)} \;\;\;\;\;\;
\\
a_{Q}^{\left( -\right)}=0
\end{array}\right), \;\;\;\;
\widetilde{S}=\left(\begin{array}{cc}
\widetilde{r} & \widetilde{t}'\\
\widetilde{t} & \widetilde{r}'
\end{array}\right).
\label{MyPHDTEc2_92a_b}
\end{equation}

Even though the scattering problem is completely described by the $\widetilde{S}$ matrix, it is important to notice that the closed channel amplitudes \begin{small}$b_{Q} ^{\left(-\right)}$\end{small}, \begin{small}$b_{Q}^{\left(+\right)}$\end{small} decrease exponentially as we move away from the disordered system, which is shown schematically in Fig. \ref{Theregimes}; moreover, those amplitudes do not contribute to the flux density current. For this reason, the scattering problem is usually described in terms of the well known {\em open channels} or {\em reduced scattering matrix} $S$, that relates open channel outgoing- and incoming-wave amplitudes in the asymptotic region, i.e.,

\begin{equation}
\left( \begin{array}{c}
b_{P}^{\left(-\right)}
\\
b_{P}^{\left(+\right)}
\end{array}
\right)=S
\left(
\begin{array}{c}
a_{P}^{\left(+\right)}
\\
a_{P}^{\left(-\right)}
\end{array}
\right),
\;\;\;\;\;S=
\left(
\begin{array}{cc}
r & t^{\prime}
\\
t & r^{\prime}
\end{array}
\right).
\label{ScatMat_Relat}
\end{equation}

We should notice, however, that the reduced matrix $S$ of a given realization of the microscopic disorder, is extracted from the generalized matrix $\widetilde{S}$, which in turn is calculated by combining the generalized scattering matrices $\widetilde{s}_{r} \sim \mathcal{T}_{r}$ ($r=1,2,\cdots, n$) of the individual scattering units: see Eqs. (5) and (6) in the Letter. This combination captures any possible open or closed channel transition, which are result of the multiple scattering processes inside the disordered system. Once the reduced matrix $S$ is known, the complex reflection $r_{aa_{0}}$ and transmission $t_{aa_{0}}$ coefficients as well as the corresponding reflectance $R_{aa_{0}}=\vert r_{aa_{0}} \vert^{2}$ and transmittance $T_{aa_{0}}=\vert t_{aa_{0}} \vert^{2}$, are easily calculated. This procedure is repeated numerically for each microscopic realization of the disordered system, which allows us to generate an ensemble of matrices $S$ and consequently to obtain, numerically, the ensemble averages $\langle r_{aa_{0}} \rangle$, $\langle t_{aa_{0}} \rangle$, $\langle R_{aa_{0}} \rangle$ and $\langle T_{aa_{0}} \rangle$.

The implementation of the GSM method described above, exhibits that the closed channel or evanescent waves contribute implicitly to the $S$ matrix given in Eq. \eqref{ScatMat_Relat}. For this reason the $S$ matrix of any microscopic realization of the microscopic disorder - and consequently the statistical properties associated to $S$ - depend formally on the closed channels, which in general are neglected in the theoretical studies of quasi one dimensional (Q1D) disordered systems. It is important to mention that, the GSM method also guaranties the flux conservation $S^{\dag}S=I$. This means that for any microscopic configuration of the disorder, the total transmittance $T_{a_{0}}=\sum_{a=1}^{N} T_{aa_{0}}$ and reflectance $R_{a_{0}}=\sum_{a=1}^{N} R_{aa_{0}}$ satisfy the condition $T_{a_{0}}+R_{a_{0}} =1$; consequently, the numerical results shown in the Letter are consistent with the flux conservation property.

\subsection{Complete set of numerical results}\label{Complete_Num_Results}

The numerical results shown in Fig. 2, illustrate the influence of $N^{\prime}=0,1,2,3$ closed channels in the expectation values of the transmittance $\left\langle T_{22}\right\rangle$, the reflectance $\left\langle R_{22}\right\rangle $, and the complex transmission $\left\langle t_{22}\right\rangle$ and reflection $\left\langle r_{22}\right\rangle$ coefficients, when the waveguide supports $N=2$ open channels. In this section we present the complete set of numerical expectation values and some complementary technical details.

\begin{figure}[t]
\begin{center}
\includegraphics[scale=0.33]{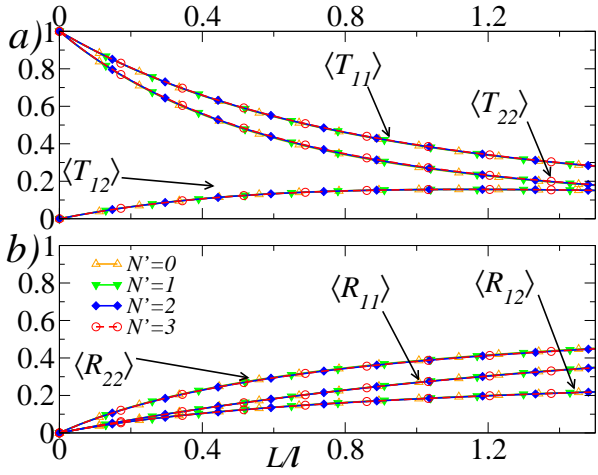}
\caption{\small{(Color online) Numerical results for $\left\langle T_{aa_{0}}\right\rangle $ and $\left\langle R_{aa_{0}}\right\rangle $ $vs$ $L/\ell$. Each simulation considers $N=2$ propagating modes and different evanescent modes ($N^{\prime}=0,1,2,3$). The results for $\left\langle T_{22}\right\rangle $ and $\left\langle R_{22}\right\rangle $ are shown in Fig. 2 of the Letter.}}
\label{Ind_Taa0_Raa0} 
\end{center}
\end{figure}

\subsubsection{Transmittance, Reflectance, Conductance and the Flux Conservation Property}

\begin{figure}[t]
\begin{center}
\includegraphics[scale=0.33]{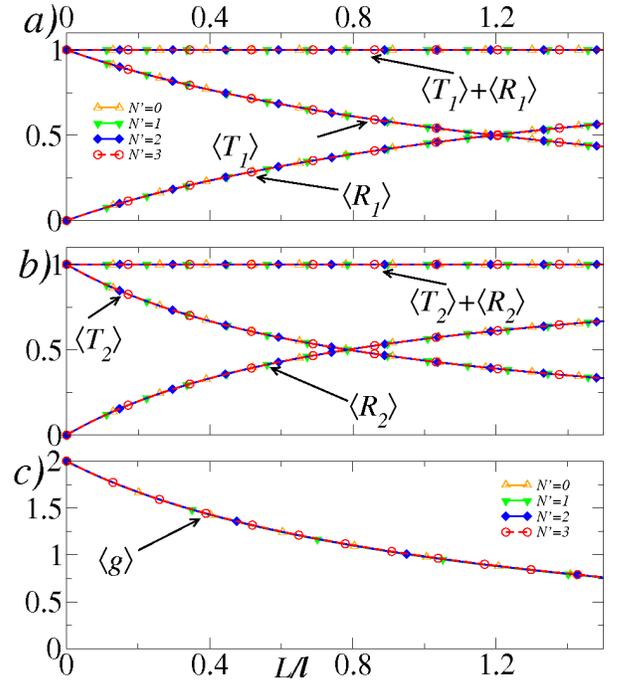}
\caption{\small{(Color online) $a$) and b) Numerical expectation values of the total transmittance $\left\langle T_{a_{0}}\right\rangle $ and reflectance $\left\langle R_{a_{0}}\right\rangle $; the flux conservation property $\left\langle T_{a_{0}}\right\rangle + \left\langle R_{a_{0}}\right\rangle =1$ is also shown. c) Numerical expectation value of the dimensionless conductance $\left\langle g \right\rangle $. Each simulation considers $N=2$ propagating modes and different evanescent modes ($N^{\prime}=0,1,2,3$).}}
\label{FluxConseProp} 
\end{center}
\end{figure}

Here in Fig. \ref{Ind_Taa0_Raa0}, we present the complete set of numerical expectation values of the transmittance $\left\langle T_{aa_{0}}\right\rangle$ and reflectance $\left\langle R_{aa_{0}}\right\rangle $. That figure exhibits the same behavior reported in the Letter: the inclusion of the closed channels does not have an important effect in $\left\langle T_{aa_{0}}\right\rangle$ and $\left\langle R_{aa_{0}}\right\rangle$. In a similar way, Fig. \ref{FluxConseProp} shows that the expectation values of the total transmittance $\left\langle T_{a_{0}} \right\rangle = \sum_{a=1}^{N} \left\langle T_{aa_{0}} \right\rangle $ and reflectance $\left\langle R_{a_{0}} \right\rangle =\sum_{a=1}^{N} \left\langle R_{aa_{0}} \right\rangle$ as well as the dimensionless conductance $\left\langle g \right\rangle = \sum_{a,a_{0}=1}^{N} \left\langle T_{aa_{0}} \right\rangle = \sum_{a_{0}=1}^{N} \left\langle T_{a_{0}}\right\rangle $ are also insensitive to the closed channel contributions. In Fig. \ref{FluxConseProp}, we can also appreciate that the numerical expectation values $\langle T_{a_{0}} \rangle$ and $\langle R_{a_{0}}\rangle$ satisfy the flux conservation property, i.e., $\langle T_{a_{0}} \rangle + \langle R_{a_{0}} \rangle=1$; therefore, the numerical results are consistent with the flux conservation property.

The numerical evidence shown in Figs. \ref{Ind_Taa0_Raa0} and \ref{FluxConseProp}, confirms previous numerical results, where the GSM method was implemented \cite{Froufe:2007}: the closed channels do not contribute in the statistics of transport coefficients. This evidence justifies the omission of the closed channels in the description of the statistical properties of the transport coefficients $T_{aa_{0}}$ and $R_{aa_{0}}$, which is a common approximation in the theoretical study of wave transport through disordered systems; however, there is not any theoretical explanation for this numerical evidence.

\subsubsection{Complex Transmission and Reflection Coefficients}

\begin{figure}[b]
\begin{center}
\includegraphics[scale=0.315]{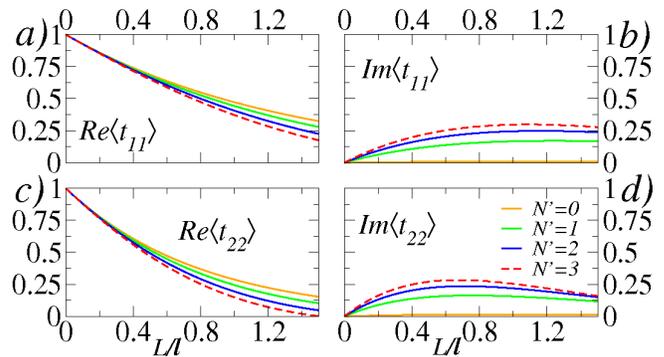}
\caption{\small{(Color online) Numerical results for $\left\langle t_{a_{0}a_{0}}\right\rangle $ $vs$ $L/\ell$. Each simulation considers $N=2$ propagating modes and different evanescent modes ($N^{\prime}=0,1,2,3$). The results for $\left\langle t_{22}\right\rangle $ is shown in Fig. 2 of the Letter.}}
\label{ScatteAmplit1} 
\end{center}
\end{figure}

The numerical results of Fig. \ref{ScatteAmplit1} show a decreasing behavior with $L/\ell$ for $\mathrm{Re}\left\langle t_{a_{0}a_{0}}\right\rangle $, which becomes more notorious as the number of closed channels ($N^{\prime}=0,1,2,3$) considered in the calculations is increased. This figure also shows that the influence of the closed channels in $\mathrm{Im}\left\langle t_{a_{0}a_{0}}\right\rangle $ is even more dramatic: if closed channels are not included in the numerical simulations ($N^{\prime}=0$), $\mathrm{Im}\left\langle t_{a_{0}a_{0}}\right\rangle $ is small ($\sim 10^{-2}$), but not strictly zero, however, when the closed channels are considered ($N^{\prime}=1,2,3$), $\mathrm{Im}\left\langle t_{a_{0}a_{0}}\right\rangle $ increases one order of magnitude.

\begin{figure}[t]
\begin{center}
\includegraphics[scale=0.30]{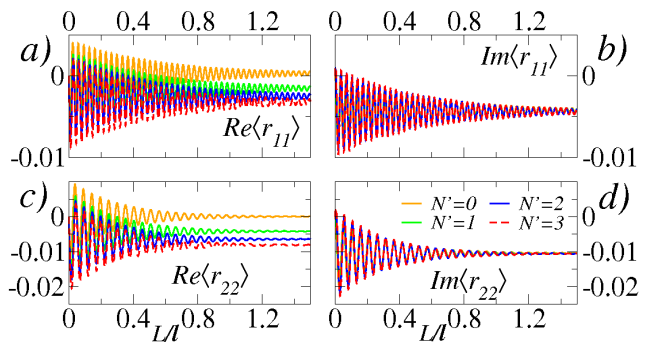}
\caption{\small{(Color online) Numerical results for $\left\langle r_{a_{0}a_{0}}\right\rangle $ $vs$ $L/\ell$. Each simulation considers $N=2$ propagating modes and different evanescent modes ($N^{\prime}=0,1,2,3$). The results for $\left\langle r_{22} \right\rangle $ are shown in Fig. 2 of the Letter.}}
\label{ScatteAmplit2}
\end{center}
\end{figure}

In Fig. \ref{ScatteAmplit2} we can appreciate a remarkable oscillatory behavior for $\left\langle r_{a_{0}a_{0}}\right\rangle $, which is rapidly attenuated as $L/\ell$ increases, regardless the number of closed channels that are considered in the calculation. The phase and amplitude of $\left\langle r_{a_{0}a_{0}} \right\rangle $ depend little on the closed channels contributions. However, the most notorious dependence on the closed channel contributions is exhibited by $\mathrm{Re}\left\langle r_{a_{0}a_{0}} \right\rangle $, whose oscillations are given around a ``background'' that depends on the number of closed channels considered in the calculations; in contrast, the four simulations of $\mathrm{Im}\left\langle r_{a_{0}a_{0}}\right\rangle $ seem to oscillates around the same ``background'', no matter how many closed channels were used in the calculations.

Previous theoretical models cannot describe the numerical evidence shown in Figs. \ref{ScatteAmplit1} and \ref{ScatteAmplit2} for $\left\langle t_{a_{0}a_{0}}\right\rangle $ and $\left\langle r_{a_{0}a_{0}}\right\rangle $, what is due to the absence of closed channels in those descriptions. As an example, we consider MT prediction \cite{PRB46:1992}, for the expectation values of the complex coefficients

\begin{eqnarray}
\left\langle t _{aa_{0}} \right\rangle^{\mathrm{(MT)}} &=& \delta_{aa_{0}} e^{-L/\ell_{a_{0}}},
\label{MyPHDEq7.89}
\\
\left\langle r _{aa_{0}} \right\rangle_{L}^{\mathrm{(MT)}} &=& 0,
\label{MyPHDEq5.33c} 
\end{eqnarray}

\noindent where $\ell_{a_{0}}$ is the scattering mean free path of the incoming open channel $a_{0}$: see Eq. (14) in the Letter.

Since in the weak scattering limit, the scattering mean free path $\ell_{a_{0}}$ does not depend on the closed channels inclusion [see Eq. (16) in the Letter], then Eq. \eqref{MyPHDEq7.89} cannot describe the influence of the closed channels in $\left\langle t_{a_{0}a_{0}} \right\rangle$, that is shown in Fig. \ref{ScatteAmplit1}. We have verified that the MT prediction $\left\langle t _{a_{0}a_{0}} \right\rangle ^{\mathrm{(MT)}} = e^{-L/\ell_{a_{0}}}$ is indistinguishable from the numerical result $\mathrm{Re}\left\langle t_{a_{0}a_{0}} \right\rangle $, when the closed channels are not considered in the calculation. However, once the closed channels are considered ($N^{\prime}=1,2,3$), $\mathrm{Re} \left\langle t _{a_{0}a_{0}} \right\rangle $ decreases faster than the MT prediction Eq. \eqref{MyPHDEq7.89}. In addition, the behavior $\mathrm{Im} \left\langle t _{a_{0}a_{0}} \right\rangle \neq 0$ is not predicted by MT result. On the other hand, although the numerical results of Fig. \ref{ScatteAmplit2} show that the amplitude of the oscillation of $\left\langle r_{a_{0}a_{0}} \right\rangle $ is small ($\sim 10^{-2}$), this expectation value is not zero, even when the closed channels are not included; therefore, the MT prediction $\left\langle r _{a_{0}a_{0}} \right\rangle^{\mathrm{(MT)}}=0$ is not able to describe the numerical results of Fig. \ref{ScatteAmplit2}.

\subsection{Theoretical results}\label{Borns_Details}

In the Letter, the Born series predictions for expectation values are use to understand the role of the closed channels in the expectation values $\langle T_{aa_{0}}\rangle $, $\langle R_{aa_{0}}\rangle $, $\langle t_{aa_{0}}\rangle $ and $\langle r_{aa_{0}}\rangle $. In the ballistic regime ($L/\ell\ll 1$) and in the SWLA ($k\ell\gg 1$), this perturbative approach shows an excellent agreement with the numerical simulations presented for $\langle T_{22}\rangle$, $\langle R_{22}\rangle$, $\langle t_{22}\rangle$ and $\langle r_{22}\rangle $: see Eqs. (18)-(19) and Fig. 2 in the Letter. In this section we give some complementary details above Born series prediction and also present its comparison with the complete set of numerical expectation values. To do this it is convenient to rewrite Eqs. (18) and (19) of the Letter as follows

\begin{widetext}
\begin{subequations}
\begin{eqnarray}
\mathrm{Re}\langle t_{aa_{0}} \rangle 
&=& \delta_{aa_{0}}
\left[1-\frac{L}{\ell_{a_{0}}}
+\left(\frac{1}{\ell_{a_{0}}^{2}}-\frac{1}{\ell_{a_{0}}^{\prime 2}}\right)\frac{L^{2}}{2!} + \cdots \right]
+O\left(\frac{1}{k\ell}\right) 
\label{ExpliRetaa0}
\\
\mathrm{Im}\langle t_{aa_{0}} \rangle 
&=& \delta_{aa_{0}}
\left[\frac{L}{\ell_{a_{0}}^{\prime}}
-\frac{L^{2}}{\ell_{a_{0}}\ell_{a_{0}}^{\prime }}  + \cdots  \right]
+O\left(\frac{1}{k\ell}\right) 
\label{ExpliImtaa0}
\\
\mathrm{Re}\langle r_{aa_{0}} \rangle 
&=& \delta_{aa_{0}}
\biggl[
\frac{-1}{2k_{a_{0}}\ell_{a_{0}}^{\prime}}
+
\left(
\frac{\cos 2k_{a_{0}}L }{2k_{a_{0}}\ell_{a_{0}}^{\prime}} - \frac{\sin 2k_{a_{0}}L }{2k_{a_{0}}\ell_{a_{0}}}
\right)
\left(1-\frac{L}{\ell_{a_{0}}}-\frac{L}{2\ell_{a_{0}a_{0}}}\right) -
\left(
\frac{\cos 2 k_{a_{0}}L}{k_{a_{0}}\ell_{a_{0}}} 
+\frac{\sin 2 k_{a_{0}}L}{k_{a_{0}}\ell_{a_{0}}^{\prime}}
\right) \frac{L}{\ell_{a_{0}}^{\prime}} +\cdots
\biggr]
\nonumber
\\
&+&O\left(\frac{1}{k\ell}\right)^{2}
\label{ExpliReraa0}
\\
\mathrm{Im}\langle r_{aa_{0}} \rangle 
&=& \delta_{aa_{0}}
\biggl[
\frac{-1}{2k_{a_{0}}\ell_{a_{0}}}
+
\left(\frac{\cos 2k_{a_{0}}L }{2k_{a_{0}}\ell_{a_{0}}} + \frac{\sin 2k_{a_{0}}L }{2k_{a_{0}}\ell_{a_{0}}^{\prime}} \right)
\left(1-\frac{L}{\ell_{a_{0}}}-\frac{L}{2\ell_{a_{0}a_{0}}}\right)
+
\left(
\frac{\cos 2 k_{a_{0}}L}{k_{a_{0}}\ell_{a_{0}}^{\prime}} 
-\frac{\sin 2 k_{a_{0}}L}{k_{a_{0}}\ell_{a_{0}}}
\right) \frac{L}{\ell_{a_{0}}^{\prime}} +\cdots
\biggr]
\nonumber
\\
&+&O\left(\frac{1}{k\ell}\right)^{2}
\label{ExpliImraa0}
\\
\left\langle
T_{aa_{0}}\right\rangle &=&
\delta_{aa_{0}}\left[1-2\frac{L}{\ell
_{a}}+2\left(\frac{L}{\ell_{a}}\right)^{2}+\cdots\right] +
\left[\frac{L}{\ell
_{aa_{0}}}+\sum_{b=1}^{N}\frac{L^{2}}{\ell_{ab}\ell_{ba_{0}}}
-\left(\frac{1}{\ell_{a}}+\frac{1}{\ell_{a_{0}}}\right)\frac{L^2}{\ell_{aa_{0}}}+\cdots\right]
+ O\left(\frac{1}{k\ell}\right)
\label{MyPHDEq7.137}
\\
\left\langle R_{aa_{0}}
\right\rangle &=&\biggl[\frac{L}{\ell_{aa_{0}}}-\biggl(\frac{1}{\ell_{a}}+\frac{1}{\ell_{aa_{0}}}+\frac{1}{\ell_{a_{0}}}
\biggr)\frac{L^2}{\ell_{aa_{0}}}
+\sum_{b=1}^{N}\frac{1+\delta_{a a_0}}{\ell_{ab}\ell_{ba_{0}}}L^2
+\cdots\biggr]+O\left(\frac{1}{k\ell}\right).
\label{MyPHDEq7.138}
\\
\left\langle g
\right\rangle &=&\sum_{a,a_{0}=1}^{N}\left\langle
T_{aa_{0}}\right\rangle = N\left(1-\frac{L}{\ell} +\frac{1}{N}\sum_{a_{0}}^{N}\left( \frac{L}{\ell_{a_{0}}}\right)^{2} +\cdots   \right) 
\end{eqnarray}
\label{Exptaa0raa0Taa0Raa0}
\end{subequations}
\end{widetext}

\noindent where the real and imaginary parts of $\langle t_{a_{0}a_{0}} \rangle$ and $\langle r_{a_{0}a_{0}} \rangle$ have been written explicitly. 

As it is discussed in the Letter, the role of the closed channels in the statistical parameters $\ell_{aa_{0}}$, $\ell_{a_{0}}$ and $\ell_{a_{0}}^{\prime}$, allows us to understand theoretically the numerical results shown in Figs. \ref{Ind_Taa0_Raa0}-\ref{ScatteAmplit2}:

\begin{itemize}

\item The expectation values of the transport coefficients $\left\langle T_{aa_{0}} \right\rangle$ and $\left\langle R_{aa_{0}} \right\rangle$ are insensitive to the closed channels inclusion, what is due to the absence of $\ell_{a_{0}}^{\prime}$ in their lowest order contributions in powers of $1/k\ell$.

\item The dominant contributions in powers of $1/k\ell$ of expectation values for the expectation values $\left\langle t_{aa_{0}} \right\rangle$ and $\left\langle r_{aa_{0}} \right\rangle$ depend strongly on the closed channels through $\ell_{a_{0}}^{\prime}$.

\end{itemize}

\subsubsection{Complex Transmission and Reflection Coefficients}

In Eq. \eqref{ExpliRetaa0}, we can appreciate that the linear term $L/\ell_{a_{0}}$ of $\mathrm{Re}\langle t_{a_{0}a_{0}} \rangle$, depends only on the scattering mean free path $\ell_{a_{0}}$, so that the first closed channel contributions appear in the quadratic term $(L/\ell_{a_{0}}^{\prime})^{2}$. This fact explains qualitatively the numerical behavior shown in Fgs. \ref{ScatteAmplit1}$a$ and \ref{ScatteAmplit1}c, where the four curves ($ N^{\prime} = 0,1,2,3 $ closed channnels) for $\mathrm{Re}\langle t_{a_{0}a_{0}} \rangle $ are quite similar for small values of $L/\ell$. On the other hand, the theoretical prediction given in Eq. \eqref{ExpliImtaa0} shows that, at the origin, the slope of $\mathrm{Im} \langle t_{a_{0}a_{0}} \rangle$ depends strongly on the closed channels through the $\ell_{a_{0}}^{\prime}$. This dependence on $\ell_{a_{0}}^{\prime}$ explains qualitatively, in the ballistic regime ($L\ll \ell$), the numerical evidence shown in Fgs. \ref{ScatteAmplit1}b and \ref{ScatteAmplit1}d for $\mathrm{Im}\langle t_{a_{0}a_{0}} \rangle $. 

The theoretical results given in Eqs. \eqref{ExpliRetaa0} and \eqref{ExpliImtaa0} suggest an exponential behavior $\left\langle t_{aa_{0}}\right\rangle \sim \delta_{aa_{0}} e^{iL/\ell_{a_{0}}^{\prime}} e^{-L/\ell_{a_{0}}}$, which reproduce the MT prediction, Eq. \eqref{MyPHDEq7.89}, when the characteristic length $\ell_{a_{0}}^{\prime}$ does not appear in the description, i.e., omitting the closed channels.

\begin{figure}[t]
\begin{center}
\includegraphics[scale=0.33]{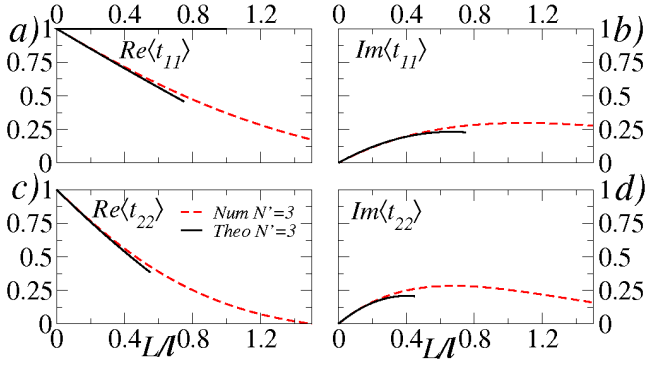}
\caption{\small{(Color online) Comparison between the numerical simulation (red dash line) and the theoretical prediction (black continuous line) for $\left\langle t_{a_{0}a_{0}}\right\rangle $. The comparison is done when the waveguide supports $N=2$ open channels and $N^{\prime}=3$ closed channels are considered. The comparison for $\left\langle t_{22}\right\rangle $ is shown in Fig. 3 of the Letter.}}
\label{NumBorntaa0} 
\end{center}
\end{figure}

In Fig. \ref{NumBorntaa0} the Born series prediction for $\left\langle t _{a_{0}a_{0}} \right\rangle$, Eqs. \eqref{ExpliRetaa0} and \eqref{ExpliImtaa0}, it is compared with the numerical simulations shown in Fig. \ref{ScatteAmplit1}. The comparison is done when $N^{\prime}=3$ closed channels are taken into account in the calculations: the agreement is excellent in the ballistic regime.

Equations \eqref{ExpliReraa0} and \eqref{ExpliImraa0} show that, both $\mathrm{Re} \langle r_{a_{0}a_{0}} \rangle$ and $\mathrm{Im} \langle r_{a_{0}a_{0}} \rangle$ have an oscillatory behavior with $e^{2k_{a_{0}}L}$, whose amplitudes are modulated by factors in powers of $1/k_{a_{0}}\ell_{a_{0}}$, $1/k_{a_{0}}\ell_{a_{0}}^{\prime}$ and in powers of $L/\ell_{a_{0}}$, $L/\ell_{a_{0}}^{\prime}$ and $L/\ell_{a_{0}a_{0}}$. To understand the numerical results shown in Fig \ref{ScatteAmplit2}, it is important to notice that, $\mathrm{Re}\langle r_{a_{0}a_{0}} \rangle$ and $\mathrm{Im}\langle r_{a_{0}a_{0}} \rangle$ contain a length independent contribution: $-1/2k_{a_{0}}\ell_{a_{0}}^{\prime}$ and $-1/2k_{a_{0}}\ell_{a_{0}}$, respectively. As we know, the mean value of the sine and cosine functions is zero, so that the theoretical curves of $\mathrm{Re}\langle r_{a_{0}a_{0}} \rangle$ and $\mathrm{Im}\langle r_{a_{0}a_{0}} \rangle$ shall oscillate, in the ballistic regime, around their corresponding $L$-independent contribution. Since $\ell_{a_{0}}^{\prime}$ is extremely sensitive to the number of closed channels considered in its evaluation [see Eq. 16 in the Letter], $\mathrm{Re}\langle r_{a_{0}a_{0}} \rangle$ suffers a displacement as we increase the number of closed channels. In contrast,  $\ell_{a_{0}}$ is insensitive to the closed channels contributions, so that $\mathrm{Im}\langle r_{a_{0}a_{0}} \rangle$ does not show any displacement with the inclusion of the closed channels. This analysis, allows to understand qualitatively, in the ballistic regime, the numerical evidence shown in Fig. \ref{ScatteAmplit2} for $\left\langle r_{a_{0}a_{0}} \right\rangle $.

\begin{figure}[b]
\begin{center}
\includegraphics[scale=0.325]{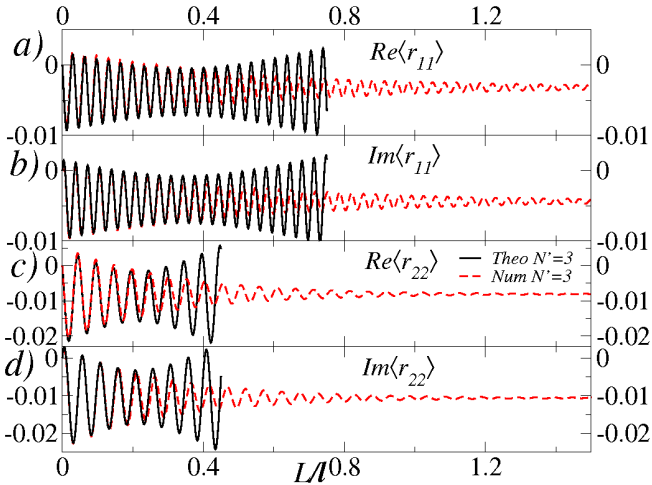}
\caption{\small{(Color online) Comparison between the numerical simulation (red dash line) and the theoretical prediction (black continuous line) for $\left\langle r_{a_{0}a_{0}}\right\rangle$. The comparison is done when the waveguide supports $N=2$ open channels and $N^{\prime}=3$ closed channels are considered. The comparison for $\left\langle r_{22}\right\rangle $ is shown in Fig. 3 of the Letter.}}
\label{NumBornraa0} 
\end{center}
\end{figure}

Figure \ref{NumBornraa0} shows the comparison between the Born series prediction for $\left\langle r _{a_{0}a_{0}} \right\rangle$, Eqs. \eqref{ExpliReraa0} and \eqref{ExpliImraa0}, and the numerical simulations of Fig. \ref{ScatteAmplit2}. The comparison is done when $N^{\prime}=3$ closed channels are taken into account in the calculations. The agreement is good enough in the ballistic regime, where the amplitude and the phase of oscillation show quantitative agreement. However, as the $L/\ell$ increases, the agreement in the phase ceases to be good, suggesting that the oscillation not only depends on $e^{2k_{a_{0}}L}$.

\subsubsection{Transmittance and Reflectance}

Equations \eqref{MyPHDEq7.137} and \eqref{MyPHDEq7.138} show the perturbative results for the expectation values $\left\langle T_{aa_{0}} \right \rangle $ and $\left\langle R_{aa_{0}} \right \rangle $. It is easy to verify, up to the order $(L/\ell)^{2}$, that those perturbative expressions are consistent with the flux conservation property, i.e.,

\begin{equation}
\langle T_{a_{0}} \rangle + \langle R_{a_{0}} \rangle=\biggl[ 1+O\left(\frac{L}{\ell}\right)^{3} \biggr] +O\biggl( \frac{1}{k\ell} \biggr).
\end{equation}

As it is mentioned in the Letter, the dominant contributions in powers of $1/k\ell$ of the expectation values $\left\langle T_{aa_{0}} \right \rangle $ and $\left\langle R_{aa_{0}} \right \rangle $ only depend on the channel-channel $\ell_{aa_{0}}$ and scattering mean free $\ell_{a_{0}}$ paths, which are insensitive to the closed channels inclusion. The closed channels contribute to the expectation values $\left\langle T_{aa_{0}} \right \rangle $ and $\left\langle R_{aa_{0}} \right \rangle $ at the order $1/k\ell$ or higher, where characteristic length $\ell_{a_{0}}^{\prime}$ appears; for instance, the first contribution in Born series expansion of $\left\langle R_{aa_{0}} \right \rangle $, is given in the following way:

\begin{equation}
\left\langle R_{aa_{0}}\right\rangle_{L}^{\mathrm{(1st\;Born)}}= \frac{L}{\ell_{aa_{0}}}+\delta_{aa_{0}} 
\left( \frac{\sin ^{2} k_{a} L}{\left( k_{a}\ell_{a}\right)^{2}} 
+ \frac{\sin ^{2} k_{a} L}{\left( k_{a}\ell_{a}^{\prime}\right)^{2}} \right).
\label{Raa0_kl_expansion}
\end{equation}

\noindent The second term of the last equation is of the order $\left( 1/k\ell\right)^{2}$, so that, in the SWLA ($k\ell\gg 1$) and in the ballistic regime ($L/\ell\ll 1$), the linear term $L/\ell_{aa_{0}}$ is the leading term of the first Born approximation; therefore, the closed channel contributions do not significantly modify the behavior of the expectation values of the transport coefficients $\left\langle T_{aa_{0}} \right \rangle $ and $\left\langle R_{aa_{0}} \right \rangle $. This explains qualitatively, at least in the ballistic regime, the numerical evidence shown in Fig. \ref{Ind_Taa0_Raa0}.

\begin{figure}[t]
\begin{center}
\includegraphics[scale=0.33]{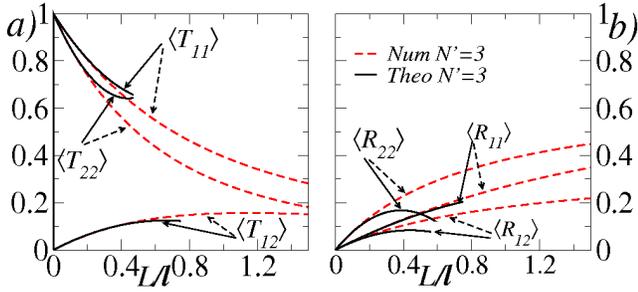}
\caption{\small{(Color online) Comparison between the numerical simulation (red dash line) and the theoretical prediction (black continuous line) for $\left\langle T_{aa_{0}}\right\rangle$ and $\left\langle R_{aa_{0}}\right\rangle$. The comparison is done when the waveguide supports $N=2$ open channels and $N^{\prime}=3$ closed channels are considered. The comparisons for $\left\langle T_{22}\right\rangle $ and $\left\langle R_{22}\right\rangle $ are shown in Fig. 3 of the Letter.}}
\label{NumBornTaa0Raa0} 
\end{center}
\end{figure}

In Fig. \ref{NumBornTaa0Raa0} we compare the Born series prediction of $\left\langle T_{aa_{0}} \right \rangle $ and $\left\langle R_{aa_{0}} \right \rangle $, Eqs. \eqref{MyPHDEq7.137} and \eqref{MyPHDEq7.138}, with the numerical results given in Fig. \ref{Ind_Taa0_Raa0}. The comparison is done when the waveguide supports $N=2$ open channels and $N^{\prime}=3$ closed channels are taken into account in the calculations. Once again, the Born series prediction gives an excellent agreement with the numerical simulations in the ballistic regime $L\ll\ell$.

\subsection{Coherent and diffuse fields}\label{Coherent_Diffsuive_Fields}

\subsubsection{Definitions}

For a given realization of the microscopic disorder the complex coefficients of the transmitted and reflected waves can be written as the sum of the average $\left\langle t_{aa_{0}}\right\rangle$, $\left\langle r_{aa_{0}}\right\rangle$ (coherent) and residual $\Delta t_{aa_{0}}$, $\Delta r_{aa_{0}}$ (diffuse) fields, i.e.,

\begin{subequations}
\begin{eqnarray}
t_{aa_{0}}=\left\langle t_{aa_{0}}\right\rangle + \Delta t_{aa_{0}}, \;\;\;\; \langle \Delta t_{aa_{0}} \rangle \equiv 0,
\\
r_{aa_{0}}=\left\langle r_{aa_{0}}\right\rangle + \Delta r_{aa_{0}}, \;\;\;\; \langle \Delta r_{aa_{0}} \rangle \equiv 0.
\end{eqnarray}
\label{Def_taa0_raa0}
\end{subequations}

\noindent $\Delta t_{aa_{0}}$, $\Delta r_{aa_{0}}$ give the statistical fluctuations around the coherent fields $\langle t_{aa_{0}} \rangle$, $\langle r_{aa_{0}} \rangle$, respectively. In a similar way, the transmittance and reflectance of a given realization are written in the following way:

\begin{subequations}
\begin{align}
T_{aa_{0}} &= \vert t_{aa_{0}}\vert^{2} =\left\langle T_{aa_{0}}\right\rangle + \Delta T_{aa_{0}},
\;\;\; \langle \Delta T_{aa_{0}} \rangle \equiv 0,
\\
R_{aa_{0}} &= \vert r_{aa_{0}}\vert^{2} = \left\langle R_{aa_{0}}\right\rangle + \Delta R_{aa_{0}},
\; \langle \Delta R_{aa_{0}} \rangle \equiv 0,
\end{align}
\label{Def_taa0taa0_raa0raa0}
\end{subequations}

\noindent where 

\begin{subequations}
\begin{eqnarray}
\langle T_{aa_{0}} \rangle &=& 
\vert\langle t_{aa_{0}} \rangle\vert^{2} + \langle \vert \Delta t_{aa_{0}}\vert
^{2}\rangle,
\label{Taa0Coehe_Diffuse_fiels}
\\
\langle R_{aa_{0}} \rangle &=& 
\vert\langle r_{aa_{0}} \rangle\vert^{2} + \langle \vert \Delta r_{aa_{0}}\vert
^{2}\rangle .
\label{Raa0Coehe_Diffuse_fiels}
\end{eqnarray}
\label{AverTaa0vsModAvertaa0}
\end{subequations}

\noindent denote, respectively, the expectation values of the transmittance and reflectance coefficients, while

\begin{eqnarray}
\Delta T_{aa_{0}} &=& \vert \Delta t_{aa_{0}} \vert ^{2} -\left\langle \vert \Delta t_{aa_{0}}\vert ^{2}\right\rangle + 2\mathrm{Re}\left( \left\langle t_{aa_{0}}\right\rangle \Delta t_{aa_{0}}^{\ast} \right) \nonumber
\\
\Delta R_{aa_{0}} &=&  \vert \Delta r_{aa_{0}} \vert ^{2} -\left\langle \vert \Delta r_{aa_{0}}\vert ^{2}\right\rangle +2\mathrm{Re}\left( \left\langle r_{aa_{0}}\right\rangle \Delta r_{aa_{0}}^{\ast} \right) \nonumber
\\
\end{eqnarray}

\noindent give the corresponding statistical fluctuations.

\subsubsection{Influence of the diffuse fields in the expectation values of the transport coefficients}

Equation \eqref{AverTaa0vsModAvertaa0} shows that the ensemble average of the transmittance $\left\langle T_{aa_{0}} \right\rangle$ and reflectance $\left\langle R_{aa_{0}} \right\rangle$ are constituted of two different contributions: the coherent field intensities $\vert\langle t_{aa_{0}} \rangle\vert^{2}$, $\vert\langle r_{aa_{0}} \rangle\vert^{2}$ and the diffusive field contributions $\langle\vert\Delta t_{aa_{0}}\vert ^{2}\rangle$, $\langle\vert\Delta r_{aa_{0}}\vert ^{2}\rangle$. In this section we briefly analyze the contributions of those kind of contributions.

The structure given in Eq. \eqref{AverTaa0vsModAvertaa0}, is numerically illustrated in Figs. 2$a$-b) of the Letter, where the ensemble average of the transport coefficients $\langle T_{22} \rangle$, $\langle R_{22} \rangle$ and the corresponding coherent intensities $ \vert\langle t_{22} \rangle\vert^{2}$, $\vert \langle r_{22} \rangle \vert^{2}$ are plotted: the difference between $\langle T_{22} \rangle$, $\langle R_{22} \rangle$ and the corresponding coherent intensities exhibit the influence of the diffusive fields $\langle \vert \Delta t_{22} \vert ^{2}\rangle$, $\langle \vert \Delta r_{22} \vert ^{2}\rangle$. In the Letter, Fig. 2$a$) shows that the coherent field intensity $ \vert\langle t_{22} \rangle\vert^{2}$ becomes less important for large values of $L/\ell$, where the diffuse field governs the behavior of $\langle T_{22} \rangle $. In contrast, Fig. 2b) shows that the diffuse field $\langle \vert \Delta r_{22}\vert ^{2}\rangle^{\mathrm{(Num)}}$ dominates the behavior of $\langle  R_{22} \rangle^{\mathrm{(Num)}}$ $\forall L/\ell$.

In order to understand the numerical evidence shown in Figs. 2$a$-b) of the Letter, in this we use the Born series prediction to identify the coherent and diffuse fields contributions.

From Eqs. \eqref{ExpliRetaa0} and \eqref{ExpliImtaa0}, it is easy to obtain, up to the order $(L/\ell)^{2}$, the coherent field contributions to the expectation value of the transmittance $\langle T_{aa_{0}} \rangle$, i.e.,

\begin{subequations}
\begin{equation}
\vert\langle t_{aa_{0}} \rangle\vert^{2}=\delta_{aa_{0}} \biggl[1-2\frac{L}{\ell_{a}}+2 \biggl(\frac{L}{\ell_{a}} \biggl)^{2}+\cdots \biggr] + O\left(\frac{1}{k\ell}\right)
\label{Coherent_field_Taa0}
\end{equation}

\noindent Since the last expression is exactly the same to the first square parenthesis in Eq. \eqref{MyPHDEq7.137}, then we can identify the second term of Eq. \eqref{MyPHDEq7.137}, as the diffusive field contributions to $\langle T_{aa_{0}} \rangle$, i.e.,

\begin{eqnarray}
\langle \vert \Delta t_{aa_{0}} \vert^{2} \rangle&=&
\biggl[
\frac{L}{\ell_{aa_{0}}}-\left(\frac{1}{\ell_{a}}+\frac{1}{\ell_{a_{0}}}\right)\frac{L^2}{\ell_{aa_{0}}}
\nonumber
\\
&+& \sum_{b=1}^{N}\frac{L^{2}}{\ell_{ab}\ell_{ba_{0}}}+\cdots 
\biggr]
+O\left(\frac{1}{k\ell}\right).
\label{Diffuse_field_Taa0}
\end{eqnarray}
\label{Coherebt_Diffuse_field_Taa0}
\end{subequations}

A similar procedure allows to demonstrate that the coherent field contribution of the reflectance $\langle R_{aa_{0}} \rangle$ is of the order $(1/k\ell)^{2}$, i.e.,

\begin{subequations}
\begin{equation}
\vert\langle r_{aa_{0}} \rangle\vert^{2}= O\left(\frac{1}{k\ell}\right)^{2}
\label{Coherent_field_Raa0}
\end{equation}

\noindent while the diffusive field contribution is given by

\begin{align}
\langle \vert \Delta r_{aa_{0}} \vert^{2} \rangle&=
\biggl[\frac{L}{\ell_{aa_{0}}}-\biggl(\frac{1}{\ell_{a}}+\frac{1}{\ell_{aa_{0}}}+\frac{1}{\ell_{a_{0}}}
\biggr)\frac{L^2}{\ell_{aa_{0}}}
\nonumber
\\
&+\sum_{b=1}^{N}\frac{1+\delta_{a a_0}}{\ell_{ab}\ell_{ba_{0}}}L^2
+\cdots\biggr]+ O\left(\frac{1}{k\ell}\right).
\label{Diffuse_field_Raa0}
\end{align}
\label{Coherebt_Diffuse_field_Raa0}
\end{subequations}

Equations \eqref{Coherebt_Diffuse_field_Taa0} and \eqref{Coherebt_Diffuse_field_Raa0} show that expectation value of the off diagonal coefficients $\langle T_{a a_{0}} \rangle$ ($a\neq a_{0}$) and each $\langle R_{a a_{0}} \rangle$ are dominated by their corresponding diffuse fields. The coherent fields are only relevant for the expectation value of the diagonal transmission coefficient $\langle T_{a_{0}a_{0}} \rangle$. The numerical results for $\langle T_{22} \rangle$ and $\langle R_{22} \rangle$ shown in Figs. 2a-b) of the Letter, are in good agreement with the last theoretical prediction.


\end{document}